\documentclass[a4paper]{article}
\usepackage{amsmath}
\begin{document}

\title{\textbf{A Note on the Painlev\'{e} Property of Coupled KdV Equations}}

\author{\textsc{Sergei Sakovich}\bigskip \\
\small Institute of Physics, Academy of Sciences, Minsk, Belarus $\diamond$ saks@tut.by}

\date{}

\maketitle

\begin{abstract}
We prove that one system of coupled KdV equations, claimed by Hirota, Hu, and Tang to pass the Painlev\'{e} test for integrability, actually fails the test at the highest resonance of the generic branch and therefore must be non-integrable.
\end{abstract}

\section{Introduction}

In Section 6 of their paper \cite{HHT}, Hirota, Hu, and Tang reported that the system of coupled KdV equations
\begin{equation}
\begin{gathered}
\frac{\partial u_i}{\partial t} + 6a \left( \sum_{k=1}^N u_k \right) \frac{\partial u_i}{\partial x} + 6(1-a) \left( \sum_{k=1}^N \frac{\partial u_k}{\partial x} \right) u_i + \frac{\partial^3 u_i}{\partial x^3} = 0 , \\[6pt]
i = 1,2, \dotsc , N , \qquad N \geq 2 ,
\end{gathered}
\label{e1}
\end{equation}
passes the Painlev\'{e} test for integrability if and only if the parameter $a$ is equal to $1$, or $1/2$, or $3/2$. The authors of \cite{HHT} pointed out that the cases $a=1$ and $a=1/2$ of \eqref{e1} correspond to integrable systems of coupled KdV equations, whereas the problem of integrability of \eqref{e1} with $a=3/2$ remains open.

In the present short note, we show that the system \eqref{e1} with $a=3/2$ actually does not pass the Painlev\'{e} test, and its integrability should not be expected therefore.

\section{Singularity analysis}

First of all, let us notice that the $N$-component system \eqref{e1} can be studied in the form of the following triangular system of two coupled KdV equations:
\begin{equation}
v_t + 6 v v_x + v_{xxx} = 0, \qquad w_t + 6a v w_x + 6(1-a) w v_x + w_{xxx} = 0,
\label{e2}
\end{equation}
where the new dependent variable $v$ is defined by $v = \sum_{k=1}^N u_k$, the new dependent variable $w$ is any one of the $N$ components $u_1 , \dotsc , u_N$, and the subscripts $x$ and $t$ denote partial derivatives. Indeed, the system \eqref{e1} is equivalent to the system consisting of the first equation of \eqref{e2} along with $N-1$ copies of the second equation of \eqref{e2} with $w = u_1 , \dotsc , u_{N-1}$ (say). Therefore, in order to check whether or not the system \eqref{e1} passes the Painlev\'{e} test, it is sufficient to consider the two equations in \eqref{e2} and do not repeat the same calculations for the remaining $N-2$ copies of the second equation of \eqref{e2}.

Setting $a=3/2$ in \eqref{e2} and starting the Weiss--Kruskal algorithm of singularity analysis \cite{WTC,JKM}, we use the expansions $v = v_0 (t) \phi^{\alpha} + \dotsb + v_r (t) \phi^{r + \alpha} + \dotsb$ and $w = w_0 (t) \phi^{\beta} + \dotsb + w_r (t) \phi^{r + \beta} + \dotsb$ with $\phi_x (x,t) = 1$, and determine branches (i.e. admissible choices of $\alpha , \beta , v_0 , w_0$) together with corresponding positions $r$ of resonances (where
arbitrary functions of $t$ can enter the expansions). The exponents $\alpha$ and $\beta$ and the positions of resonances turn out to be integer in all branches. In what follows, we only consider the generic singular branch, where $\alpha = \beta = -2$, $v_0 = -2$, $w_0 (t)$ is arbitrary, and $r=-1,0,1,4,6,8$. This branch describes the singular behavior of generic solutions. There are also two non-generic branches, but they correspond to the constraints $w_0 =0$ and $w_0 = w_1 = 0$ imposed on the generic branch and do not require any separate consideration therefore.

Substituting the expansions
\begin{equation}
v = \sum_{n=0}^{\infty} v_n (t) \phi^{n-2} , \qquad w = \sum_{n=0}^{\infty} w_n (t)  \phi^{n-2}
\label{e3}
\end{equation}
with $\phi_x (x,t) = 1$ into the system \eqref{e2} with $a=3/2$, we obtain the following recursion relations for the coefficients $v_n$ and $w_n$ of \eqref{e3}:
\begin{equation}
\begin{gathered}
(n-2)(n-3)(n-4) v_n + 3 (n-4) \sum_{i=0}^n v_i v_{n-i} + \\
(n-4) \phi_t v_{n-2} + v_{n-3,t} = 0 , \\[6pt]
(n-2)(n-3)(n-4) w_n + 3 \sum_{i=0}^n (3n-4i-4) v_i w_{n-i} + \\
(n-4) \phi_t w_{n-2} + w_{n-3,t} = 0 , \\[6pt]
n = 0,1,2,3, \dotsc ,
\end{gathered}
\label{e4}
\end{equation}
where the subscript $t$ denotes the derivative with respect to $t$, and $v_k = w_k = 0$ for $k = -3, -2, -1$ formally.

Now we have to check whether the recursion relations \eqref{e4} are compatible at the resonances.

The resonance $-1$, as always, corresponds to the arbitrariness of the function $\psi$ in $\phi = x + \psi (t)$.

We have $v_0 = -2$ in \eqref{e4} at $n=0$, for the chosen branch. The function $w_0 (t)$ remains arbitrary, which corresponds to the resonance 0.

Setting $n=1$ in \eqref{e4}, we find that $v_1 = 0$, while the function $w_1 (t)$ remains arbitrary, and the compatibility condition at the resonance 1 is satisfied.

At $n=2$ and $n=3$, which are not resonances, we get from \eqref{e4}, respectively,
\begin{equation}
v_2 = - \frac{1}{6} \phi_t , \qquad w_2 = \frac{1}{12} w_0 \phi_t
\end{equation}
and
\begin{equation}
v_3 = 0 , \qquad w_3 = \frac{1}{60} w_1 \phi_t + \frac{1}{30} w_{0,t} .
\end{equation}

Setting $n=4$ in \eqref{e4}, we find that
\begin{equation}
w_4 = - \frac{1}{2} v_4 w_0 + \frac{1}{48} w_{1,t} ,
\end{equation}
while the function $v_4 (t)$ remains arbitrary, and the compatibility condition at the resonance 4 is satisfied.

At $n=5$, which is not a resonance, we find from \eqref{e4} that
\begin{equation}
\begin{gathered}
v_5 = - \frac{1}{36} \phi_{tt} , \\[6pt]
w_5 = - \frac{1}{4} v_4 w_1 - \frac{1}{7200} w_1 \phi_t^2 + \frac{1}{900} w_{0,t} \phi_t + \frac{1}{72} w_0 \phi_{tt} .
\end{gathered}
\end{equation}

Setting $n=6$ in \eqref{e4}, we obtain
\begin{equation}
w_6 = - \frac{1}{2} v_6 w_0 - \frac{1}{14400} w_{1,t} \phi_t + \frac{31}{3600}w_1 \phi_{tt} + \frac{1}{1800} w_{0,tt} ,
\end{equation}
while the function $v_6 (t)$ remains arbitrary, and the compatibility condition at the resonance 6 is satisfied.

At $n=7$, which is not a resonance, we get from \eqref{e4} the following:
\begin{equation}
\begin{gathered}
v_7 = - \frac{1}{24} v_{4,t} , \\[6pt]
w_7 = - \frac{1}{2} v_6 w_1 + \frac{17}{1680} v_4 w_1 \phi_t + \frac{1}{201600} w_1 \phi_t^3 + \frac{1}{48} v_{4,t} w_0 - \\
\frac{1}{105} v_4 w_{0,t} - \frac{1}{25200} w_{0,t} \phi_t^2 + \frac{1}{2016} w_{1,tt} .
\end{gathered}
\end{equation}

The highest resonance in the chosen branch is 8. Setting $n=8$ in the recursion relations \eqref{e4}, we find that
\begin{equation}
v_8 = - \frac{1}{6} v_4^2 + \frac{1}{2592} \phi_{ttt} ,
\end{equation}
the function $w_8 (t)$ remains arbitrary, but the compatibility condition at the resonance 8 is not satisfied, and we obtain the following constraint imposed on some of arbitrary functions appeared at lower resonances:
\begin{equation}
300 v_{4,t} w_1 - 7 w_1 \phi_t \phi_{tt} + 6 w_{0,t} \phi_{tt} = 0 .
\label{e12}
\end{equation}

The appearance of the constraint \eqref{e12} means that the Laurent type expansions \eqref{e3} do not represent the general solution of the studied system, and we have to modify the expansion for $w$ by introducing logarithmic terms, starting from the term proportional to $\phi^6 \log \phi$. This non-dominant logarithmic branching of solutions is a clear symptom of non-integrability. Consequently, the case $a=3/2$ of the system \eqref{e2}---and of the system \eqref{e1}, equivalently---fails the Painlev\'{e} test.

\section{Conclusion}

We have shown that, contrary to the claim of Hirota, Hu, and Tang \cite{HHT}, the system of coupled KdV equations \eqref{e1} with $a=3/2$ does not pass the Painlev\'{e} test for integrability.

Let us note, moreover, that the singularity analysis of coupled KdV equations has been addressed in the papers \cite{Kar} and \cite{Sak}, published prior to \cite{HHT}. In particular, the integrable cases $a=1$ and $a=1/2$ of the system \eqref{e1} can be found in \cite{Sak} as the systems (vi) and (vii), respectively, which have passed the Painlev\'{e} test, whereas the case $r_1 = 1$ in Section 2.1.3 of \cite{Sak} predicts that the system \eqref{e1} with $a=3/2$ must fail the Painlev\'{e} test for integrability.

The obtained result that the system of coupled KdV equations \eqref{e1} with $a=3/2$ actually does not pass the Painlev\'{e} test for integrability explains very well why no Lax representation has been proposed as yet for this case of coupled KdV equations.

\section*{Acknowledgment}

The author is grateful to the anonymous reviewer for valuable comments.

\end{document}